%% file: article-arxiv.tex
\documentclass[12pt,letterpape]{article}
\usepackage[a4paper, total={7in, 10in}]{geometry}

\usepackage{graphicx}
\usepackage{helvet}
\usepackage{authblk}
\usepackage{hyperref}
\usepackage{amsmath} 
\usepackage{amssymb} 
\usepackage{orcidlink} 
\usepackage[super,comma,sort&compress]  
   {natbib}

\usepackage[T1]{fontenc}    
\usepackage{hyperref}       
\usepackage{url}            
\usepackage{booktabs}       
\usepackage{amsfonts}       
\usepackage{nicefrac}       
\usepackage{microtype}      
\usepackage{lipsum}
\usepackage[table]{xcolor}
\usepackage{comment}
\usepackage{float}

\newcommand{\norm}[1]{\left \lVert #1 \right \rVert}

\input{TeX_definitions}

\makeatletter
\renewcommand{\maketitle}{\bgroup\setlength{\parindent}{0pt}
\begin{flushleft}
  \textbf{\@title}
  
  \@author
\end{flushleft}\egroup}
\makeatother

\title{The Reward Function and the Least Cost Principle for Gravitation and other Laws of Physics}
\date{}

\vspace{1cm}
 
\author[1-3,*]{Rubén Moreno-Bote}

\affil[1]{Department of Engineering, Universitat Pompeu Fabra, 08002, Barcelona, Spain}
\affil[2]{Center for Brain and Cognition, Universitat Pompeu Fabra, 08002, Barcelona, Spain}
\affil[3]{Serra Húnter Fellow Programme, Universitat Pompeu Fabra, Barcelona, 08002, Spain}

\affil[*]{Correspondence: ruben.moreno@upf.edu}

\begin{document}


\maketitle

\subsection*{Abstract}
If the universe follows a specific design, then a central question is which cost function is optimized by the observed forces.
This is the problem of inverse optimal control, or inverse reinforcement learning, in which a reward function is inferred from the dynamics of the observed system.
We first establish the {\em least cost principle}, whereby the laws of motion can be derived from minimization of a time-discounted integral of the acceleration cost minus a state-dependent reward function.
After determining the functional form of the acceleration cost from basic principles, we infer the reward function from the laws of motion governing classical gravitation and Coulomb forces. 
The inferred reward function is high when pairs of particles have high relative velocities and when their relative motion is orthogonal to their distance vectors. 
All in all, our work suggests that relative motion and quasi-circular orbits are the dynamical and static features optimized by central forces in nature.

\newpage

If the universe has an intelligent design, a central question is which cost function is minimized.
The cost function—or, more naturally, the reward function—could endow the universe with "desired" features that natural forces ought to optimize. 
A natural objective might be one that produces rich and structured motion. 
However, such motion should arise with minimal intervention by the interaction forces. 
The form of the cost and reward functions, even for classical interactions such as Newtonian gravitation or Coulomb forces, has not yet been established.

Here, we use inverse optimal control (IOC) \cite{kalman1964linear,moylan1973nonlinear,casti1980general,azar2020inverse} to derive the reward function from the system's dynamics.
These and related methods in inverse reinforcement learning have been used to infer cost functions in both linear \cite{kalman1964linear,wu2026output,xue2021inverse} and nonlinear \cite{moylan1973nonlinear,ziebart2008maximum,xu2022energy} control problems, including impulse control \cite{wang2025optimal}, and are widely applied to infer reward functions from observed animal behavior \cite{drugowitsch2012cost,rothkopf2013modular}. 
However, IOC is limited by the complexity of the adopted control cost, as well as by assumptions about time horizon and time discounting. 

In this paper, we establish the {\em least cost principle}. This principle implies that the laws of motion in classical systems with central forces, such as Newtonian gravitation and Coulomb forces, arise as the optimal solution to a control problem \cite{kirk2004optimal,fleming2006controlled} that minimizes the time-discounted integral of acceleration cost minus a reward function.
The least cost principle resembles the {\em least action principle} \cite{landau1960mechanics}, but the former is an optimal control problem, whereas the latter is an optimal path problem \cite{kirk2004optimal}. 
Both principles lead to identical laws of motion, much like uncountably many different parabolas share the same minimum; their individual merits lie in the distinct insights that each framework provides.

From basic principles—rather than being assumed for mathematical convenience \cite{kalman1964linear,moylan1973nonlinear,fleming2006controlled}—we show 
that the control cost associated with the least cost principle is quadratic, taking the form of a sum of squared accelerations across particles. 
Applying IOC using the least cost principle and the derived acceleration cost, we obtain the form of the unknown reward function for Newtonian gravitation and Coulomb forces.
We find that the inferred reward function promotes two salient features: it is positive for high relative motion between pairs of particles and negative when their relative motion is not orthogonal to their distance vectors. 
These effects are inversely modulated by the distance between the particles.

All in all, we uncover two salient features that natural forces optimize: relative motion and circular-like orbits. 
Optimizing these features may provide the universe with fundamental ingredients for the emergence of complexity.
The maximization of motion and its diversity is also a central idea in other principles, such as the maximum occupancy principle \cite{moreno2023empowerment,ramirez2024complex} and empowerment \cite{klyubin2005empowerment}. In this paper, we show that motion emerges as an optimized quantity in nature, rather than being assumed as a principle. 
Our quantitative characterization of the reward function also goes well beyond the qualitative description of gravitation and other central forces as producing "attraction" or "repulsion" between pairs of particles, by demonstrating that the generation of relative motion and circular-like orbits is actively promoted.


{\em The Least Cost Principle} —
We define the time-discounted cumulative cost of a trajectory $x_{ik}(t)$, $t \geq t_0$, of $i=1,\cdots,N$ particles with Cartesian coordinates $k$ as
\begin{equation}
  \mathcal{C} = \int_{t_0}^{\infty} dt \; e^{-\gamma(t-t_0)} 
            \left[ \sum_{i} \frac{1}{2} m_i \norm{\ddot{x}_{i}(t)}^2 - R(x(t),\dot{x}(t)) \right]    
            \; ,
    \label{eq:cost}
\end{equation}
with fixed initial coordinates $x_0=x(t_0)$ and velocities $\dot{x}_0=\dot{x}(t_0)$, where we use the
notation $x=(x_1,\cdots,x_N)$, with $x_i=(x_{i1},x_{i2},x_{i3})$.
Here, $\gamma>0$ is the temporal discount exponent and the term $C_a(\ddot{x}) = \sum_{i} \frac{1}{2} m_i \norm{\ddot{x}_{i}(t)}^2$ is the acceleration (control) cost.
The state reward function $R(x,\dot{x})$ depends on the system state, which includes the particle coordinates $x$ and their velocities $\dot{x}$. Importantly, the reward does not explicitly depend on accelerations $\ddot{x}$; the only dependence on accelerations appears in the acceleration cost.
The acceleration cost implements a penalty for having large accelerations, such that strong forces are discouraged. 
We will refer to $\mathcal{C}$ as simply cost-to-go, and assume that it is finite.


The cost-to-go $\mathcal{C}$ is a functional of the trajectories.
The equations of motion are determined by finding the acceleration trajectory $\ddot{x}^*(t)$, $t \geq t_0$, that minimize the cost-to-go with fixed initial conditions $x_0$ and $\dot{x}_0$, 
\begin{equation}
     \mathcal{C}^*(x_0,\dot{x}_0) = \min_{\ddot{x}} \mathcal{C} , \; \; \ddot{x}^* =   \arg\min_{\ddot{x}} \mathcal{C}   
            \; ,
    \label{eq:minimum_cost}
\end{equation}
where $\mathcal{C}^*$ is the minimum (optimal) cost-to-go; by definition it is time-invariant and thus independent of $t_0$. 

Eqs. \ref{eq:cost}-\ref{eq:minimum_cost} define the {\em least cost principle}. 
Thus, obtaining the laws of motion reduces to an optimal control problem \cite{kirk2004optimal,fleming2006controlled,lewis2009reinforcement}. 
This problem effectively amounts to maximizing the reward function—which contains the desired dynamical and static features—while using the weakest possible forces.
The least cost principle is related to, but different from the least action principle \cite{landau1960mechanics}.
While the action is a functional of the Lagrangian, similar to a cost function, it does not explicitly depend on accelerations, in contrast to the cost-to-go in Eq. \ref{eq:cost}. 

The form of the acceleration cost in Eq. \ref{eq:cost} is uniquely determined by symmetry principles and invariances (see details in Appendix \ref{sec:appendix_acceleration_cost}).
We first impose the condition that the acceleration cost is time and rotational invariant, and that it is additive over particles and their masses. 
Second, we impose that changes in the acceleration cost induced by internal forces are invariant to the choice of a homogeneously accelerated reference frame. 
These two sets of conditions are sufficient to determine the acceleration cost $C_a(\ddot{x})$ identically as in Eq. \ref{eq:cost}.
Quadratic cost terms are typically assumed in optimal control problems for mathematical convenience \cite{kirk2004optimal,fleming2006controlled,lewis2009reinforcement}. Here we have shown that this form arises from first principles in closed systems. 

We note that when the reward function is zero, the cost-to-go reduces to 

\noindent
$\mathcal{C} = \int_{0}^{\infty} dt e^{-\gamma t} \sum_{i} \frac{1}{2} m_i \norm{\ddot{x}_{i}(t)}^2$. 
Minimizing this cost-to-go using Eq. \ref{eq:minimum_cost} gives rise to the equations of motion $\ddot{x}_{ik}=0$. Their solution corresponds to rectilinear motion at constant speed for each of the $N$ particles. Therefore, this optimal control problem corresponds to $N$ independent particles.

{\em Inverse Optimal Control for General Forces} —
Our goal is to determine the unknown reward function in Eq. \ref{eq:cost} given the laws of motion. We show that inferring the reward function in a time-discounted cost-to-go for general forces is tractable. 

The optimal cost-to-go in Eq. \ref{eq:cost} can be expressed in the form of a Bellman equation by heuristically dividing time into differential steps $dt$, expanding the exponential and the next-step optimal cost up to first order in $dt$, $dx$, and $d\dot{x}$ as 
\begin{align}
    \mathcal{C}^* &= \min_{\ddot{x}} 
        \Bigg\{ 
        \left( \sum_{ik} \frac{1}{2} m_i \ddot{x}^2_{ik} - R(x,\dot{x}) \right)dt
        \nonumber
        \\
        + & (1-\gamma dt) \left( \mathcal{C}^* + \sum_{ik} \frac{\partial \mathcal{C}^*}{\partial x_{ik}} dx_{ik} 
        + \sum_{ik} \frac{\partial \mathcal{C}^*}{\partial \dot{x}_{ik}} d\dot{x}_{ik} \right)
        \Bigg\}
         \; ,
    \label{eq:bellman_eq}
\end{align}
where $\mathcal{C}^*=\mathcal{C}^*(x,\dot{x})$.
After using the identities $dx=\dot{x}dt$ and $d\dot{x}=\ddot{x}dt$, minimization over the accelerations leads to the optimal equations of motion 
\begin{equation}
    m_i \ddot{x}_{ik} = - \frac{\partial \mathcal{C}^*}{\partial \dot{x}_{ik}}
    \; ,
    \label{eq:equations_of_motion}
\end{equation}
and to the Hamilton-Jacobi-Bellman equation, which explicitly expresses the unknown reward function as a function of the optimal cost-to-go as 
\begin{equation}
    R(x,\dot{x})  
       = 
       \sum_{ik} 
       \left[ \frac{1}{2 m_i} \left(  \frac{\partial \mathcal{C}^*}{ \partial \dot{x}_{ik}}   \right)^2
       + \frac{\partial \mathcal{C}^*}{\partial x_{ik}} \dot{x}_{ik} 
        + \frac{\partial \mathcal{C}^*}{\partial \dot{x}_{ik}} \ddot{x}_{ik}
        \right]
        - \gamma \mathcal{C}^*
         .
    \label{eq:bellman_eq_solution}
\end{equation}


Given known forces $F_{ik}(x)$, from $m_i \ddot{x}_{ik} = F_{ik}$ and Eq. \ref{eq:equations_of_motion} the optimal cost-to-go becomes
\begin{equation}
    \mathcal{C}^*(x,\dot{x}) = -\sum_{ik} F_{ik}(x) \dot{x}_{ik} + B(x)
    \; ,
    \label{eq:equations_cost_force}
\end{equation}
where $B(x)$ is an arbitrary function independent of the velocities $\dot{x}$. 
For $B(x)=0$, the optimal cost-to-go is the negative time derivative of the work, $\frac{dW}{dt}=- \sum_{ik} F_{ik}(x) \dot{x}_{ik}$. 
Generalization to the case of velocity-dependent forces is straightforward and explicit if a primitive can be found. 
After inserting the solution of Eq. \ref{eq:equations_cost_force} into Eq. \ref{eq:bellman_eq_solution}, the unknown reward function takes the form 
\begin{align}
 R(x,\dot{x}) & = - \sum_{ik,jp} \dot{x}_{ik} \dot{x}_{jp}  \frac{\partial F_{ik}} {\partial x_{jp}}  
    - \frac{1}{2} \sum_{ik} \frac{F^2_{ik}}{m_i} 
    \; ,
 \nonumber
    \\
 &+ \sum_{ik} \dot{x}_{ik} \left( \gamma F_{ik} + \frac{\partial B}{\partial x_{ik}} \right) - \gamma B(x)
 \label{eq:state_cost_function}
\end{align}
which is our second central result.
If the force is conservative with potential energy $U(x)$, then the first term can be rewritten in a more symmetric form as $-\sum_{ik,jp} \dot{x}_{ik} \dot{x}_{jp}  \frac{\partial^2 U }{\partial x_{ik} \partial x_{jp}}$.

As we will see below, the first term in Eq. \ref{eq:state_cost_function} contains contributions favoring high velocities. The second term clearly penalizes strong forces. 

Our derivation shows that the equations of motion $m_i \ddot{x}_{ik} = F_{ik}$ minimize the cost-to-go in Eq. \ref{eq:cost} under the reward function in Eq. \ref{eq:state_cost_function}. 
This can also be explicitly shown (see details in Appendix \ref{sec:appendix_optimal_force_fixed_state_cost_function}): trajectories generated by the dynamics $m_i \ddot{x}_{ik} = \tilde{F}_{ik}$ with $\tilde{F}_{ik}$ different from $F_{ik}$ leads to a non-lower cost-to-go in Eq. \ref{eq:cost} when the reward function is chosen as in Eq. \ref{eq:state_cost_function}.

We now specialize Eq. \ref{eq:state_cost_function} to two important cases:



{\em i) The Reward Function for Classical Gravitation} —
The total gravitational force on particle $i$ due a system of $N$ particles is
\begin{equation}
    F_{ik}(x) = - \sum_{j, \; j \neq i} \frac{G m_i m_j}{r_{ij}^3} (x_{ik}-x_{jk})
    \;,
\end{equation}
with $r_{ij} = \norm{x_i-x_j}$.
After some straightforward but lengthy calculations, the reward function in Eq. \ref{eq:state_cost_function} with $B(x)=0$ becomes
\begin{align}
    R(x,\dot{x}) & = \frac{G}{2} \sum_{i \neq j} \frac{m_i m_j}{r_{ij}^3} \norm{\dot{x}_i - \dot{x}_j}^2
    & \textit{(I)}
    \nonumber
    \\
       - &\frac{3 G}{2} \sum_{i \neq j} \frac{m_i m_j}{r_{ij}^5} 
        [ (x_i - x_j) \cdot (\dot{x}_i - \dot{x}_j)]^2
    & \textit{(II)}
     \nonumber
    \\
       - &\frac{G^2}{2} \sum_{i \neq l, i \neq s} \frac{1}{m_i} \frac{m_i^2 m_l m_s}{r^3_{il} r^3_{is}} 
        (x_i - x_l) \cdot (x_i - x_s)
    & \textit{(III)}
    \label{eq:state_cost_gravitation}
\end{align}
where we have omitted the term proportional to $\gamma$, as it can be be made arbitrarily small. 
In the last and next equations, sums run over all indexes involved. 
Masses are not canceled for later convenience.
The optimal cost-to-go is from Eq. \ref{eq:equations_cost_force}
\begin{equation}
    \mathcal{C}^*(x,\dot{x}) = \frac{G}{2} \sum_{i \neq j} \frac{m_i m_j}{r_{ij}^3} 
         (x_i - x_j) \cdot (\dot{x}_i - \dot{x}_j)
         \;.
         \label{eq:optimal_cumulative_cost}
\end{equation}
Both the reward function and the optimal cost-to-go depend only on relative positions and velocities, as expected for a closed system. 

From Eqs. \ref{eq:equations_cost_force}-\ref{eq:state_cost_function}, the optimal cost-to-go and reward function are defined up to an arbitrary function $B(x)$. As a consequence, $B(x)$ can be chosen such that an arbitrary function of $x$ alone, as well as an arbitrary term linear in $\dot{x}$, can be added to the reward function. This implies that only Terms I and II are fundamental, since they are quadratic in the velocity and therefore cannot be absorbed into the contributions of $B(x)$. 
Furthermore, Term III arises from the presence of the derived quadratic control cost in the cost-to-go, whereas Terms I and II are of a more intrinsic nature.

First, reward Term I is non-negative and depends on the norm of the difference between the velocity vectors of pairs of particles. Consequently, this reward term is large when the relative motion between particles is high. 
This effect is also modulated by the distance between particles: high relative motion is promoted at short distances, but at large distances the reward function becomes neutral.

Second, reward Term II is non-positive and depends on the dot product between the distance vector and the relative velocity vector of pairs of particles. 
A pairwise contribution to this reward becomes zero when the relative motion between the pair of particles is perpendicular to their distance vector. Consequently, Term II promotes circular-like motion of particles around one another.
Non-circular motion is not penalized at long distances but becomes costly at short distances. 


\begin{figure}[htbp]
{
  \centering 
  \includegraphics[scale=1.]{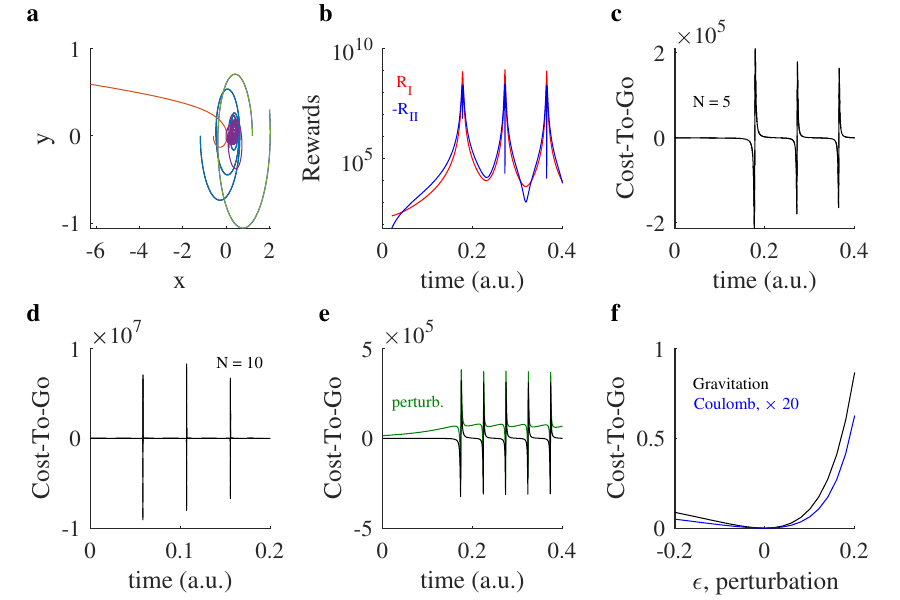}
  \caption{Newtonian gravitation maximizes the reward function in Eq. \ref{eq:state_cost_gravitation}. 
  (\textbf{a}) Trajectories of five particles following Newtonian gravitation. 
  (\textbf{b}) Reward contributions from Terms I and II in Eq. \ref{eq:state_cost_gravitation} as a function of time. Since reward Term II is non-positive, its absolute value is shown. 
  (\textbf{c}) The minimum cost-to-go in Eq. \ref{eq:optimal_cumulative_cost} for each state along the trajectory, plotted as a function of time (black line). This coincides with the actual cost incurrent by the trajectory numerically computed via Eq. \ref{eq:cost} (red line; not visible). 
  (\textbf{d}) With an increasing number of particles, $N=10$, the cost-to-go fluctuates more widely but spends more time at relatively small values.  
  (\textbf{e}) A small perturbation of the Newtonian force from $1/r^2$ to $1/r^{2+\epsilon}$ with $\epsilon=0.05$ produces a higher cost-to-go (green line) than the optimal one (black). The cost-to-go is computed at each state along the trajectories generated with the perturbed force.
  (\textbf{f}) Cost-to-go (black line) incurred by a perturbed Newtonian force $1/r^{2+\epsilon}$ as a function of $\epsilon$, calculated using the same initial condition as in panel (b). The cost-to-go attains a minimum at $\epsilon=0$. 
  Cost-to-go (blue line) incurred by a perturbed Coulomb force $1/r^{2+2\epsilon}$ as a function of $\epsilon$.
  Parameters are specified in Appendix \ref{sec:parameters}.
  }
  \label{fig_1}
  }
\end{figure}

Numerical simulations confirm our analytical results. 
As expected \cite{landau1960mechanics}, the generated trajectories are mostly curvilinear, and particles tend to form circular-like orbits when they are close to each other (Fig. \ref{fig_1}a; $N=5$ particles).
The reward contributions of Terms I and II fluctuate over time (Fig. \ref{fig_1}b), with Term I always positive and Term II always negative.
The minimum cost-to-go is relatively small for most of the trajectory but is punctuated by semi-periodic positive and negative peaks (Fig. \ref{fig_1}c). 
We verified that numerical integration of the cost-to-go (red line, not visible) matches the analytically derived cost-to-go in Eq. \ref{eq:optimal_cumulative_cost} (black). 
For a larger system with $N=10$ particles, the fluctuations in cost-to-go increase, but the cost-to-go remains small for a larger fraction of the trajectory compared to the $N=5$ particle system.
The cost-to-go increases if the particles follow a non-Newtonian force $1/r^{2+\epsilon}$ (Fig. \ref{fig_1}e).
The minimum cost-to-go is attained when $\epsilon=0$, corresponding to purely Newtonian gravitation (Fig. \ref{fig_1}f).


{\em ii) The Reward Function for Coulomb Forces} —
For Coulomb electrostatic forces, the reward function is identical to Eq. \ref{eq:state_cost_gravitation}, with $G$ replaced by $-C_{Coulomb}$ and the masses $m_i$ in the numerators replaced by the charges $q_i$, while masses in the denominators remain unchanged.
For charges of opposite sign, the same rules apply as for Terms I and II in gravitation.
However, for charges of the same sign, the features are inverted: relative motion is minimized, and the motion of particles tends to align parallel to their distance vectors, with identical intensity regardless of whether motion is toward or away from each other. 

As in the case of Newtonian gravitation, if the Coulomb force is perturbed from $1/r^2$ to $1/r^{2+2\epsilon}$, the resulting cost-to-go is higher than for the unperturbed case $\epsilon = 0$ (Fig. \ref{fig_1}f). 



{\em Discussion} —
We have inferred the reward function optimized by classical forces by first establishing the least cost principle and deriving the form of the acceleration cost from first principles. 
Our analysis has uncovered the dynamical and static features optimized by Newtonian gravitation, Coulomb interactions, and more general forces.
In particular, relative motion and circular-like orbits are favored. 
These features are arguably two essential ingredients for the design of an "interesting" universe, as structured motion forms the basis for the emergence of complexity. 

The choice of a non-zero temporal discount exponent $\gamma<1$ has several implications.
First, it allows for mathematical tractability and avoids cumbersome derivations when searching for time-invariant reward functions \cite{kalman1964linear,moylan1973nonlinear}. 
Second, a non-zero temporal discount implicitly assumes that the system can evaporate or disintegrate, which is a reasonable assumption. 
Third, assuming $\gamma<1$ emphasizes transients rather than the stationary distribution, which is appropriate if the expected lifetime of the universe is finite. 
In contrast, taking $\gamma=1$ would imply that the cost-to-go is dominated by the asymptotic stationary distribution; that is, the horizon would be so long that transient effects become negligible. 
Although this approach is sometimes adopted in reinforcement learning \cite{hazan2019provably,grytskyy2023general}, it may preclude the emergence of dynamical complexity across multiple temporal scales.  

We finally note that reward Terms I and II arise from the first term in the general expression of the reward function in Eq. \ref{eq:state_cost_function}, which is linear in the force. 
An important implication is that if the reward function optimized by each of a set of forces is known, then the reward function optimized by the sum of those forces corresponds to the sum of the individual reward functions.
This, in turn, implies that the motion and circularity features we derived remain relevant even when new forces are added to the system. 

\vspace{0.2cm}
{\em Acknowledgments} — I thank Dmytro Grytskyy for discovering typos in the equations. This project was supported by grants funded by the Spanish Ministry of Science, Innovation and Universities (MICIU/AEI/10.13039/501100011033) and by ICREA ACADÈMIA (2022) funded by the Catalan Institution for Research and Advanced Studies.

\bibliography{references}

\newpage

{\normalfont \textbf{Appendix} }  

\section{The Acceleration Cost for $N$ Independent Particles}
\label{sec:appendix_acceleration_cost}

A fundamental question is why the acceleration cost in Eq. \ref{eq:cost} has the form
\begin{equation}
    C_a(\ddot{x}) = \sum_{i} \frac{1}{2} m_i \norm{\ddot{x}_{i}}^2
    \;.
    \label{eq:acceleration_cost}
\end{equation}
For instance, powers $\norm{\ddot{x}_{i}}^n$ with $n \neq 2$ could in principle be used.
Indeed, all of them would lead to identical rectilinear equations of motion $\ddot{x}_{ik}=0$.
We next describe sufficient conditions for the derivation of the acceleration cost form in Eq. \ref{eq:acceleration_cost}.

First, because of time homogeneity, the acceleration cost cannot explicitly depend on time. 
Second, for $N$ independent particles and due to additivity, the acceleration cost should be the sum of the individual acceleration costs of unit-mass particles $C_a(\ddot{x}_i)$, 
\begin{equation}
    C_{a}(\ddot{x}) = \sum_{i} m_i C_{a}(\ddot{x}_i)
    \;,
\end{equation}
where $\ddot{x}_i$ is the acceleration vector of particle $i$.
Third, because of spatial isotropy, the unit-mass single particle acceleration cost should be a function of the squared modulus of the accelerations, that is, $C_{a}(\ddot{x}_i) = C_{a}(\norm{\ddot{x}_i}^2)$. 
We note that we are abusing notation when using the same symbol for different functions. 
We adopt the notational convention that different arguments indicate that the function can be different.

Finally, we impose invariance in the acceleration cost changes due to internal forces across different reference frames.
We consider a system of $N$ particles, and two different reference frames.
The particles are initially non interacting, and they have accelerations $\ddot{x}_i$.
These accelerations are due, for instance, to an external field such as gravity. 
Now, we assume that at some time $t$, the particles interact through mutual forces. 
This interaction causes a sudden change of accelerations from $\ddot{x}_i$ to $\ddot{x}'_i = \ddot{x}_i + b_i$, where $b_i$ is the acceleration change for particle $i$. 
We impose that the change in acceleration cost before and after the sudden force onset be invariant to the reference frame chosen. 
This is trivially guaranteed for any inertial reference frame, as the accelerations of all particles are the same regardless of the reference frames chosen.
Therefore, we ask the more stringent condition that the change in acceleration cost should be invariant to any {\em homogeneously accelerated reference frame} chosen. 
For instance, the change in acceleration cost of our system of particles should be the same within or outside a gravitational field. 
This condition suffices to fully determine the functional form of the acceleration cost, Eq. \ref{eq:acceleration_cost}, as we show next.

In the first reference frame the accelerations of the system particles change from is $\ddot{x}_i$ to $\ddot{x}'_i = \ddot{x}_i + b_i$ after the sudden onset of the internal forces. 
The change in acceleration cost is then
\begin{equation}
    \Delta C_a = \sum_i m_i ( C_a(\norm{\ddot{x}'_i}^2)  - C_a(\norm{\ddot{x}_i}^2 )
    \; .
\end{equation}

In the second reference frame there is an additional homogeneous acceleration $a(t)$ with $k$-th component $a_k(t)$; the acceleration vector is the same for all particles, and it can be time-dependent.
In this reference frame, the acceleration of particle $i$ changes from $\ddot{x}_i+a(t)$ to $\ddot{x}_i+ b_i + a(t)$ after the sudden onset of the internal forces.
The change in acceleration cost is then 
\begin{equation}
    \Delta \tilde{C}_a = \sum_i m_i ( C_a(\norm{\ddot{x}'_i + a}^2) - C_a(\norm{\ddot{x}_i + a}^2) 
    \; .
\end{equation}

Therefore, our invariance condition requires the identity $\Delta  \tilde{C}_a = \Delta C_a$, as the change in acceleration cost should be the same in any homogeneously accelerated reference frame.

Expanding the last equation up to first order in the acceleration $a$, 
\begin{equation}
    \Delta \tilde{C}_a = \Delta C_a 
    + \sum_i m_i \left( \frac{d}{dy}C_a(\norm{\ddot{x}'_i}^2) \; \ddot{x}'_i 
            - \frac{d}{dy} C_a(\norm{\ddot{x}_i}^2) \; \ddot{x}_i \right) \cdot a 
    + O(\norm{a}^2)
    \; ,
    \label{eq:invariance_accel_cost}
\end{equation}
where $\frac{d}{dy}C_a(\norm{\ddot{x}_i}^2)$ is the derivative of the real-valued function $C_a(y)$ evaluated at $\norm{\ddot{x}_i}^2$.
Invariance implies that the first order term $a$ vanishes.
This condition holds if the derivative $\frac{d}{dy}C_a(\norm{\ddot{x}_i}^2)$ is a constant, because from Newton Second Law, $\sum_i m_i b_i = 0$ in a closed system. 
Therefore, we obtain $C_a(\norm{\ddot{x}_i}^2)=\frac{1}{2} \norm{\ddot{x}_i}^2$ choosing the standard factor $\frac{1}{2}$, and thus we recover the acceleration cost for $N$ particles in Eq. \ref{eq:acceleration_cost}.

It is easy to see that the change in acceleration cost in Eq. \ref{eq:acceleration_cost} is invariant in any homogeneously accelerated reference frame not only to first order in $a(t)$, but exactly at all orders. 

Furthermore, no other choice of the (differentiable) function $C_a(y)$ satisfies invariance.
This can be seen more simply by using the case of two particles, $N=2$. 
We define the functions $g_k(\ddot{x}_i) = \frac{d}{dy}C_a(\norm{\ddot{x}_i}^2) \ddot{x}_{ik} $.
The vanishing condition at first order in Eq. \ref{eq:invariance_accel_cost} implies  
$m_1 (g_k(\ddot{x}'_1) - g_k(\ddot{x}_1)  ) + m_2 (g_k(\ddot{x}'_2) - g_k(\ddot{x}_2)  ) = 0$ for all $k$, which means that $m_2/m_1 = - (g_k(\ddot{x}'_1) - g_k(\ddot{x}_1)) /(g_k(\ddot{x}'_2) - g_k(\ddot{x}_2))$.
(The solution $g_k(\ddot{x}) = d$, with $d$ constant, is not allowed because it would lead to non differentiable cost at zero, or the trivial solution that the acceleration cost is a constant for $d=0$.)
Newton Second Law implies that $m_2/m_1 = - b_{1k}/b_{2k} = -( \ddot{x}'_{1k} - \ddot{x}_{1k}) /(\ddot{x}'_{2k} - \ddot{x}_{2k})$. 
Since the function $g_k$ should be independent of the masses, equating we obtain $(g_k(\ddot{x}'_1) - g_k(\ddot{x}_1)) /(g_k(\ddot{x}'_2) - g_k(\ddot{x}_2)) = ( \ddot{x}'_{1k} - \ddot{x}_{1k}) /(\ddot{x}'_{2k} - \ddot{x}_{2k}) $.
This condition should hold for any values of the coordinates consistent with having positive masses.
For instance, we can take for all $k$ $\ddot{x}_{1k}=c$ arbitrary, $\ddot{x}'_{1k}>c$, $\ddot{x}_{2k}=1$ and $\ddot{x}'_{2k}=0$, which implies that
$g_k(\ddot{x}'_1) = g_k(c) + (g_k(1)-g_k(0)) (\ddot{x}'_{1k} - c)$ (we denote by $c$, $1$ and $0$ the vectors with identical entries $c$, $1$ and $0$ respectively).
By definition $g_k(0)=0$, and thus $g_k(\ddot{x}'_1) = (g_k(1)-g_k(0)) \ddot{x}'_{1k}$, which means that $g_k$ is proportional to the $k$-th component of the acceleration.
Therefore, the derivative $\frac{d}{dy}C_a$ is again found to be constant, and thus we arrive to the solution in Eq. \ref{eq:acceleration_cost}.

\section{The Optimal Force for Fixed Reward Function}
\label{sec:appendix_optimal_force_fixed_state_cost_function}

Here we show that for a fixed reward function $R(x,\dot{x})$ determined by some force functions $F_{ik}(x)$ as in Eq. \ref{eq:state_cost_function}, along with some regularity conditions, the trajectory that leads to the absolute minimum of the cost-to-go in Eq \ref{eq:cost} is given by the equations of motion $\ddot{x}_{ik} = F_{ik}$.
In other words, any other force $\tilde{F}_{ik} \neq F_{ik}$ would lead to non-lower cost-to-go. 

From Eq. \ref{eq:cost}, the cost-to-go is
\begin{equation}
  \mathcal{C} = \int_{t_0}^{\infty} dt \; e^{-\gamma(t-t_0)} 
            \left[ \sum_{i} \frac{1}{2} m_i \norm{\ddot{x}_{i}(t)}^2 - R(x(t),\dot{x}(t)) \right]    
            \; ,
    \label{eq:cost_appendix}
\end{equation}
where $x$ and its temporal derivatives refer to the coordinates, velocities and accelerations of the particles, and we have taking the initial time to be zero, $t_0=0$.
The reward function associated with the equation of motion $m_i \ddot{x}_{ik} = F_{ik}(x)$ is, from Eq. \ref{eq:state_cost_function},
\begin{equation}
 R(x,\dot{x}) = - \sum_{ik,jp} \dot{x}_{ik} \dot{x}_{jp}  \frac{\partial F_{ik}} {\partial x_{jp}}  
    - \frac{1}{2} \sum_{ik} \frac{F^2_{ik}}{m_i} 
    + \gamma \sum_{ik} \dot{x}_{ik}  F_{ik}
 \label{eq:state_cost_function_appendix}
\end{equation}
where we have taken $B(x)=0$.
Therefore, inserting Eq. \ref{eq:state_cost_function_appendix} into Eq. \ref{eq:cost_appendix}, the cost-to-go for initial condition $x_0$ and $\dot{x}_0$ is
\begin{equation}
      \mathcal{C}(x_0,\dot{x}_0) = \int_{0}^{\infty} dt \; e^{-\gamma t} 
            \bigg[  \sum_{ik} \frac{1}{2} m_i \ddot{x}^2_{ik}       
            + \sum_{ik,jp} \dot{x}_{ik} \dot{x}_{jp}  \frac{\partial F_{ik}} {\partial x_{jp}}  
    + \frac{1}{2} \sum_{ik} \frac{F^2_{ik}}{m_i} 
    - \gamma \sum_{ik} \dot{x}_{ik}  F_{ik}
            \bigg]    
            \; .
    \label{eq:cost_in_appendix}
\end{equation}
The integral over the second term on the right-hand side can be rewritten integrating by parts as 
\begin{align}
      \int_{0}^{\infty} dt \; e^{-\gamma t } \sum_{ik,jp} \dot{x}_{ik} \dot{x}_{jp}  \frac{\partial F_{ik}} {\partial x_{jp}}
      & =  \int_{0}^{\infty} dt \; e^{-\gamma t} \sum_{ik} \dot{x}_{ik} \frac{d}{dt}  F_{ik}(x)
\nonumber
\\
      &=  e^{-\gamma t} \sum_{ik} \dot{x}_{ik} F_{ik}(x) \bigg|_{0}^{\infty} 
       - \int_{0}^{\infty} dt \; e^{-\gamma t}  \sum_{ik}( -\gamma \dot{x}_{ik} + \ddot{x}_{ik} ) F_{ik}(x)
\nonumber
\\
      &= - \sum_{ik} \dot{x}_{ik,0} F_{ik}(x_0)
      - \int_{0}^{\infty} dt \; e^{-\gamma t} \sum_{ik}( -\gamma \dot{x}_{ik} + \ddot{x}_{ik} ) F_{ik}(x)
            \; ,
\end{align}
where we have assumed that the $\dot{x}_{ik} F_{ik}(x)$ remains bounded and therefore the decaying exponential term dominates at long times.
Inserting this expression back into Eq. \ref{eq:cost_in_appendix}, 
\begin{align}
      \mathcal{C}(x_0,\dot{x}_0) &= - \sum_{ik} \dot{x}_{ik,0} F_{ik}(x_0)
      + \int_{0}^{\infty} dt \; e^{-\gamma t} 
            \sum_{ik} \left( \frac{1}{2} m_i \ddot{x}^2_{ik} - \ddot{x}_{ik}  F_{ik} + \frac{1}{2 m_i} F_{ik}^2 \right)    
      \nonumber
\\
      &= - \sum_{ik} \dot{x}_{ik,0} F_{ik}(x_0)
      + \int_{0}^{\infty} dt \; e^{-\gamma t} 
            \sum_{ik}  \frac{1}{2 m_i} (m_i \ddot{x}_{ik} - F_{ik})^2    
            \; .
    \label{eq:cost_1D_in2}
\end{align}
The first term in the right hand sie is precisely the minimum cost-to-go in Eq. \ref{eq:equations_cost_force}.
This means that we need to prove that when the dynamics follows $m_i \ddot{x}_{ik}=F_{ik}(x)$ the integral is zero, and when the dynamics follows $m_i \ddot{x}_{ik}=\tilde{F}_{ik}(x)$, with $\tilde{F}_{ik}(x) \neq F_{ik}(x)$, the remaining integral is non-negative. 
That the integral is zero when $m_i \ddot{x}_{ik}=F_{ik}(x)$ is obvious by direct substitution.
Also, it is clear that the integral is non-negative for any choice of the dynamics different from $m_i \ddot{x}_{ik}=F_{ik}(x)$ because masses are non-negative. 
This proves our initial statement.

\section{Parameters}
\label{sec:parameters}

In Fig. \ref{fig_1}, panels a-c, we perform simulations of 5 particles under Newtonian gravitation. They are labeled as $k=\{0,1,2,3,4\}$, where particle $k=2$ is heavy, with mass $m_1=10$, while the other particles are lighter with masses $m_2 = 1+0.01k$, $k=\{0,1,3,4\}$. We use $G=1$, $\gamma = 10$, simulation time $T=2$ (only a segment is plotted), and integration time step $\delta t = 10^{-7}$.
Initial conditions on the $x$ coordinate are $x_0 = 2(k - 2)/5$, while in the $y$ and $z$ coordinates are all zero. 
Velocities only have non-zero $y$ components, with value $v_{y,0} = 4(k - 2)/5$.

In Fig. \ref{fig_1}d, we perform simulations of 10 particles under Newtonian gravitation. They are labeled as $k=\{0,\cdots,9\}$, where particle $k=5$ is heavy, with mass $m_1=10$, while the other particles are lighter with masses $m_2 = 0.01(k+1)$, $k=\{0,\cdots,4,6,\cdots,9\}$. We use $G=1$, $\gamma = 10$, simulation time $T=1$ (only a segment is plotted), and integration time step $\delta t = 10^{-10}$.
Initial conditions on the $x$ coordinate are $x_0 = 3(k - 4.5)/10$, while in the $y$ and $z$ coordinates are all zero. 
Velocities only have non-zero $y$ coordinates, with value $v_{y,0} = 3(k - 4.5)/5$.

In Fig. \ref{fig_1}e, we perform simulations of 5 particles under perturbed Newtonian gravitation. 
All the parameters are as in Fig. \ref{fig_1}a-c, except that the Newtonian force has been perturbed from $1/r^2$ to $1/r^{2+\epsilon}$ with $\epsilon=0.05$. Trajectories are generated under the perturbed force, and the cost-to-go over each state of the trajectory is computed in two different ways. First, using the minimum cost-to-go in Eq. \ref{eq:optimal_cumulative_cost} as if for each state from then onward the force would be back to the Newtonian force. And second, the cost-to-go produced by the perturbed force is numerically computed by a line integral over the generated trajectories using Eq. \ref{eq:cost} with the reward function exactly as in Eq. \ref{eq:state_cost_gravitation} (with no perturbation added). The obtained cost-to-go produced by the perturbed force is shown to be always above the minimum cost-to-go.

In Fig. \ref{fig_1}f, we perform simulations of 3 particles under perturbed Newtonian gravitation and under Coulomb forces.
For gravitation, we choose $G=1$ and $\gamma = 5$, simulation time $T=5$, and integration time step $\delta t = 10^{-7}$.
The 3 particles are labeled as $k=\{0,1,2\}$, where particle $k=1$ is heavy, with mass $m_1=10$, while the other particles are lighter with masses $m_2 = 0.01(k+1)$, $k=\{0,2\}$.  
Initial conditions on the $x$ coordinate are $x_0 = k - 1$, while in the $y$ and $z$ coordinates are all zero. 
Velocities only have non-zero $y$ coordinates, with value $v_{y,0} = 2(k - 1)$.

For Coulomb forces, the heavy particle has mass $m_1=10$, while the other particles are lighter with identical masses $m_2 = 0.01$. The heavy mass has positive charge $q=1$, and the lighter particles have negative charge $q=-1$.
We choose $C_{Coulomb}=1$ and $\gamma = 5$, simulation time $T=5$, and integration time step $\delta t = 10^{-7}$.
Initial conditions on the $x$ coordinate are $x_0 = k - 1$, while in the $y$ and $z$ coordinates are all zero. 
Velocities only have non-zero $y$ coordinates, with value $v_{y,0} = 2(2 + 0.01k)(k - 1)$.

We did not observe any qualitative variation in the presented results when testing for different parameters, including initial conditions.

\end{document}

%% file: TeX_definitions.tex
\newcommand{\be}{\begin{equation}}
\newcommand{\ee}{\end{equation}}
\newcommand{\ba}{\begin{align}}
\newcommand{\ea}{\end{align}}
\newcommand{\bea}{\begin{eqnarray}}
\newcommand{\eea}{\end{eqnarray}}

\usepackage{amsmath}

\newfloat{Box}{tbp}{lob}
\definecolor{boxbg}{rgb}{0.9,0.9,0.9}
\newcounter{theoryboxc}
\newsavebox{\savetheorybox}
\newcommand{\printtheorybox}[1]{\fcolorbox{black}{boxbg}{#1}}
{\begin{Box}\begin{lrbox}{\savetheorybox}%
 \begin{minipage}{0.985\linewidth}
 \refstepcounter{theoryboxc}%
 \textbf{Box \thetheoryboxc. #1} %
}
{\end{minipage}%
 \end{lrbox}%
 \printtheorybox{\usebox{\savetheorybox}}%
 \end{Box}
}